\documentclass[11pt,twoside]{article}%
\usepackage{asp2010}%
\resetcounters%

\begin{document}%

\title{Observations and NLTE Modeling of the Gaseous Planetary Debris Disk around Ton\,345}%
\author{S.~Hartmann, T.~Nagel, T.~Rauch, and K.~Werner}%
\affil{Institute for Astronomy and Astrophysics, Kepler Center for Astro and Particle Physics, Eberhard Karls University, Sand 1, 72076 T{\"u}bingen, Germany}%

\begin{abstract}%
Debris disks around single white dwarfs are thought to be the remains of planetary bodies disrupted by tidal forces. Ongoing accretion of the hereby produced dust allows to detect the planetary material in the white dwarf photosphere and to conclude on its chemical composition. As an alternative, the composition can in principle be determined directly from the emission lines of the sometimes additionally observed gaseous component of the disks. To this aim, we perform spectral modeling with our non-LTE code for accretion disks. We have obtained new observations of Ton\,345 in order to look for long- and short-term variations in the disk line-profiles. We find that the prominent red-violet asymmetry of the Ca\,\textsc{ii} infrared triplet almost disappeared. Line-profile variations during one night are not seen without doubt.%
\end{abstract}%

\section{Metal-rich Gaseous Debris Disks Around White Dwarfs}%
For several years, there is an evermore growing number of single white dwarfs (WD) detected showing a high infrared (IR) flux. The picture of an evolved central object shredding and accreting parts of its surrounding planetary system \citep{Debes:2002,Jura:2003} became the prevailing explanation for this excess. The high metallicity and mass of the hereby formed dusty debris disks resulting from a destroyed asteroid or minor planet fit the observed data very well \cite[e.g.][]{Farihi:2009}. While most of the time spectral analysis of these objects focuses on the accreted material using the WD's atmosphere as a ``detector'', the discovery of emission lines \cite[e.g.][]{Gansicke:2006} points towards an additional gaseous disk component and allows to study the planetary debris directly.%

\section{Ton\,345}%
\begin{figure}[ht!]%
\centering%
\includegraphics[width=0.75\textwidth]{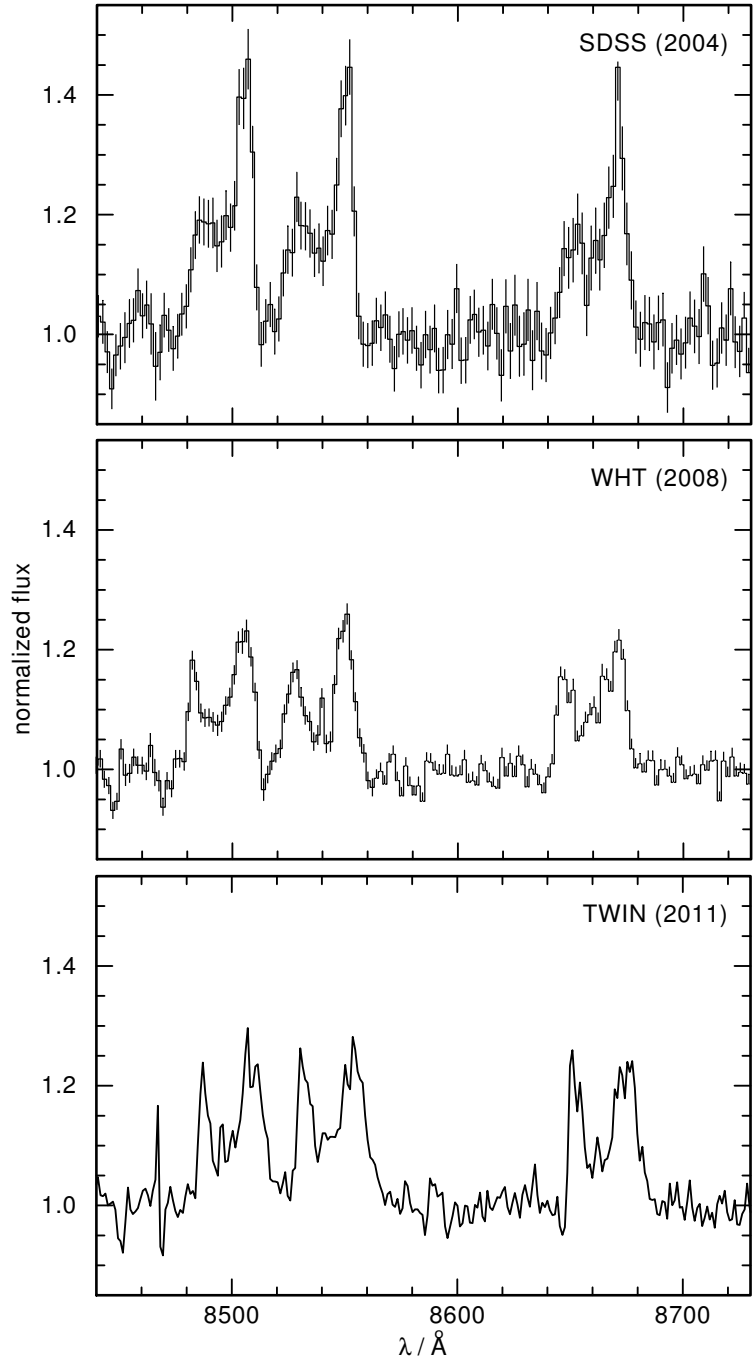}%
\caption{\label{fig:1}SDSS and WHT spectra (top and center panels; datasets taken from \citealt{Gansicke:2008}) of Ton\,345 in the range of the IRT. The co-added spectrum of the recently obtained \mbox{Calar~Alto} data is shown in the bottom panel.}
\end{figure}

Ton\,345 (WD\,0842+231) is the first DBZ-type WD ($T_{\textrm{\scriptsize eff}}=$ 18\,600\,K, $\log g=$ 8.2, \citealt{Gansicke:2008}) at which the Doppler-broadened Ca\,\textsc{ii} IR triplet (IRT) at $\lambda\lambda$ 8498, 8542, 8662\,{\AA}, the hallmark for gaseous debris disks, was discovered by \cite{Gansicke:2008}. Two spectra, taken by the Sloan Digital Sky Survey (SDSS) in early 2004 and with the William Herschel Telescope (WHT) in late 2008 (upper two panels in Fig.~\ref{fig:1}), show a strong decrease of the equivalent widths and a significant red-violet asymmetry of the double-peaked profiles.%

\begin{figure}[bth]%
\centering%
\includegraphics[angle=-90,width=\textwidth]{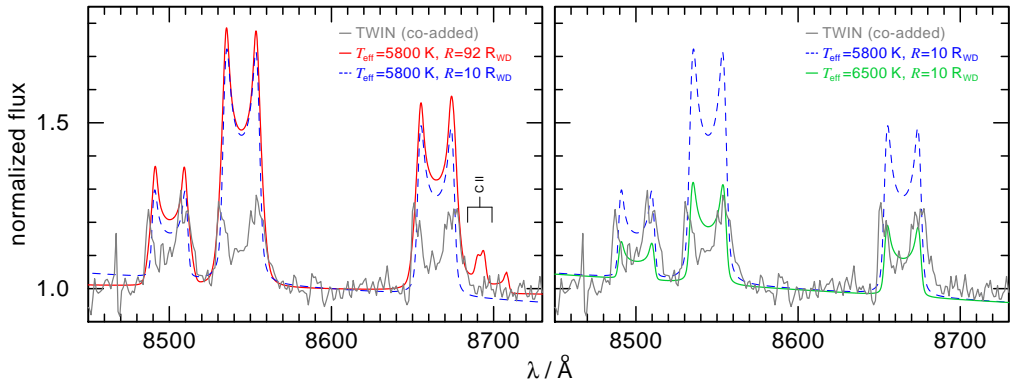}%
\caption{\label{fig:2}The IRT in the co-added TWIN spectra compared to three different non-LTE disk models. In the left panel both NLTE models have the same constant effective temperature $T_{\textrm{\scriptsize eff}}=$ 5800\,K but different radial extent. The models in the right-hand side panel differ in $T_{\textrm{\scriptsize eff}}$ while the radial extent of the disk is kept constant at $R=10\,\textrm{R}_{\textrm{\scriptsize WD}}$.}
\end{figure}%

\begin{figure}[bth]%
\centering%
\includegraphics[width=0.63\textwidth]{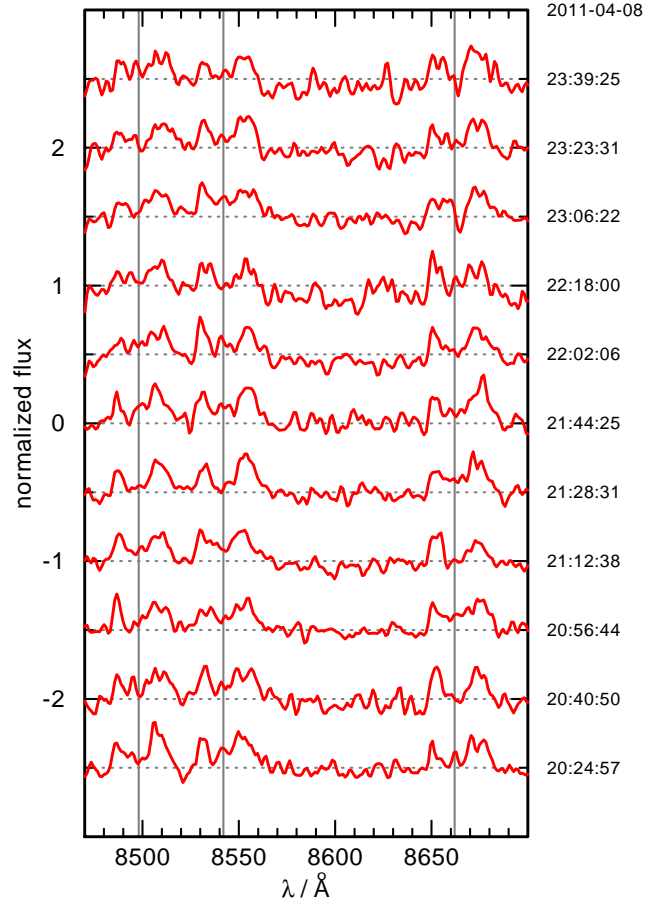}%
\caption{\label{fig:3}Normalized \mbox{Calar Alto} spectra in chronological order (bottom to top), obtained within about three hours. The vertical lines indicate the rest wavelengths of the IRT components.}%
\end{figure}%

\section{Calar Alto Observations}%
The observations were performed with the Cassegrain TWIN Spectrograph at the 3.5\,m telescope at \mbox{Calar~Alto~Observatory}, using gratings T08 for the blue  and T04 for the red channel (dispersion 72\,{\AA}/mm) to cover the wavelength ranges 3500-6500\,{\AA} and 5500-9000\,{\AA}, respectively. 17 consecutive spectra were obtained on the night of UTC\,2011-04-08, with 900\,s integration time each. As conditions were not ideal, the S/N of some spectra is rather poor. The eleven best exposures, i.e.\ $\textrm{S/N}>6$, were co-added (bottom panel in Fig.~\ref{fig:1}).\par%
To study the disk parameters, we calculated non-LTE radiation-transfer models using our code for accretion disk spectra \texttt{AcDc} \citep{Nagel:2004}. The models shown in Fig.~\ref{fig:2} represent non-stationary, metal-rich gaseous disks with different but constant effective temperature $T_{\textrm{\scriptsize eff}}(r)$ and different radial extent $R$. The observed line width is best reproduced by using a rather cool ($T_{\textrm{\scriptsize eff}}=$ 5800\,K) and widely extended ($R=10\,\textrm{R}_{\textrm{\scriptsize WD}}$) model (left-hand side panel Fig.~\ref{fig:2}, red solid line). However, to fit the similar relative line strengths of the three IRT components, a model with higher temperature ($T_{\textrm{\scriptsize eff}}=$ 6500\,K) and only $R=92\,\textrm{R}_{\textrm{\scriptsize WD}}$ is necessary (green solid line, right panel Fig.~\ref{fig:2}).%

\section{Observed Line-Profile Variations}%
The frequently observed changes in the IRT of gaseous debris disks around single WDs \citep[e.g.\ ][]{Gansicke:2008,Melis:2010} give the opportunity to study the dynamical processes occurring within the disk. First attempts to explain the lines' asymmetry by describing the disk geometry in a non-axissymmetric way based on hy\-dro\-dy\-namical models might explain the temporal changes \citep{Hartmann:2010,Hartmann:2011}.\par%
In the case of Ton\,345, we see in our observations that the long-term trend reported by \cite{Gansicke:2008} continued. The red-violet asymmetry almost disappeared (Fig.~\ref{fig:1}). Due to insufficient S/N ratio of the single spectra (Fig.~\ref{fig:3}), a short-term variability in the line-profiles can not be claimed without a doubt. Future observations with a 8\,m-class telescope are necessary for more definitive conclusions.%

\acknowledgments SH is supported by the Ger\-man Re\-search Foun\-da\-tion (DFG, grant WE\,1312/45-1). TR is supported by the Ger\-man Aero\-space Cen\-ter (DLR, grant 50\,OR\,0806).%

\bibliography{hartmann}%

\end{document}